\documentclass[%
reprint, 
nofootinbib,
 amsmath,amssymb,
 aps,
]{revtex4-1}
\usepackage{epsfig,epsf,rotating,wrapfig,tabularx}
\usepackage{dcolumn}
\usepackage{bm}
\usepackage{graphicx}

\begin{document}

\preprint{APS/123-QED}

\title{Hadroproduction in heavy-ion collisions }

\author{\firstname{A.~A.}~\surname{Bylinkin}}
 \email{alexander.bylinkin@desy.de}
\affiliation{%
 Institute for Theoretical and Experimental
Physics, ITEP, Moscow, Russia
}%
\author{\firstname{A.~A.}~\surname{Rostovtsev}}
 \email{rostov@itep.ru}
\affiliation{%
 Institute for Theoretical and Experimental
Physics, ITEP, Moscow, Russia
}%
\author{\firstname{N.~S.}~\surname{Chernyavskaya}}
\affiliation{%
 Institute for Theoretical and Experimental
Physics, ITEP, Moscow, Russia
}%

\begin{abstract}
The shapes of invariant differential cross section for charged particle production as function of transverse momentum measured in heavy-ion collisions are analyzed. 
The data measured at RHIC and LHC are treated as function of energy density according to a recent theoretical approach. The Boltzmann-like statistical distribution is extracted  from the whole statistical ensemble of produced hadrons using the introduced model. Variation of the temperature, characterizing this exponential distribution, is studied as function of energy density.
\end{abstract} 

\pacs{Valid PACS appear here}
\maketitle

\section{Introduction}
Inclusive charged particle distributions have been studied for a long time to derive the general properties of hadronic interactions at high energies. A large body of the experimental data on charge particle production spectra in baryon-baryon, gamma-baryon and gamma-gamma collisions has been accumulated during last forty years. However, the underlying dynamics of hadron production in high energy particle interactions is still not fully understood. 

Recently, a new unified approach to describe the particle production spectra shape was proposed~\cite{OUR1}. It was suggested to approximate the charged particle spectra as function of the particle's transverse momentum by a sum of an exponential (Boltzmann-like) and a power law statistical distributions:
\begin{equation}
\label{eq:exppl}
\frac{d\sigma}{P_T d P_T} = A_e\exp {(-E_{Tkin}/T_e)} +
\frac{A}{(1+\frac{P_T^2}{T^{2}\cdot N})^N},
\end{equation}
where  $E_{Tkin} = \sqrt{P_T^2 + M^2} - M$
with M equal to the produced hadron mass. $A_e, A, T_e, T, N$ are the free parameters to be determined by fit to the data.  The detailed arguments for this particular choice are given in~\cite{OUR1}.  For the charged hadron spectra a mass of hadrons is assumed to be equal to the pion mass. 



\begin{figure}[!ht]
\includegraphics[width =8cm]{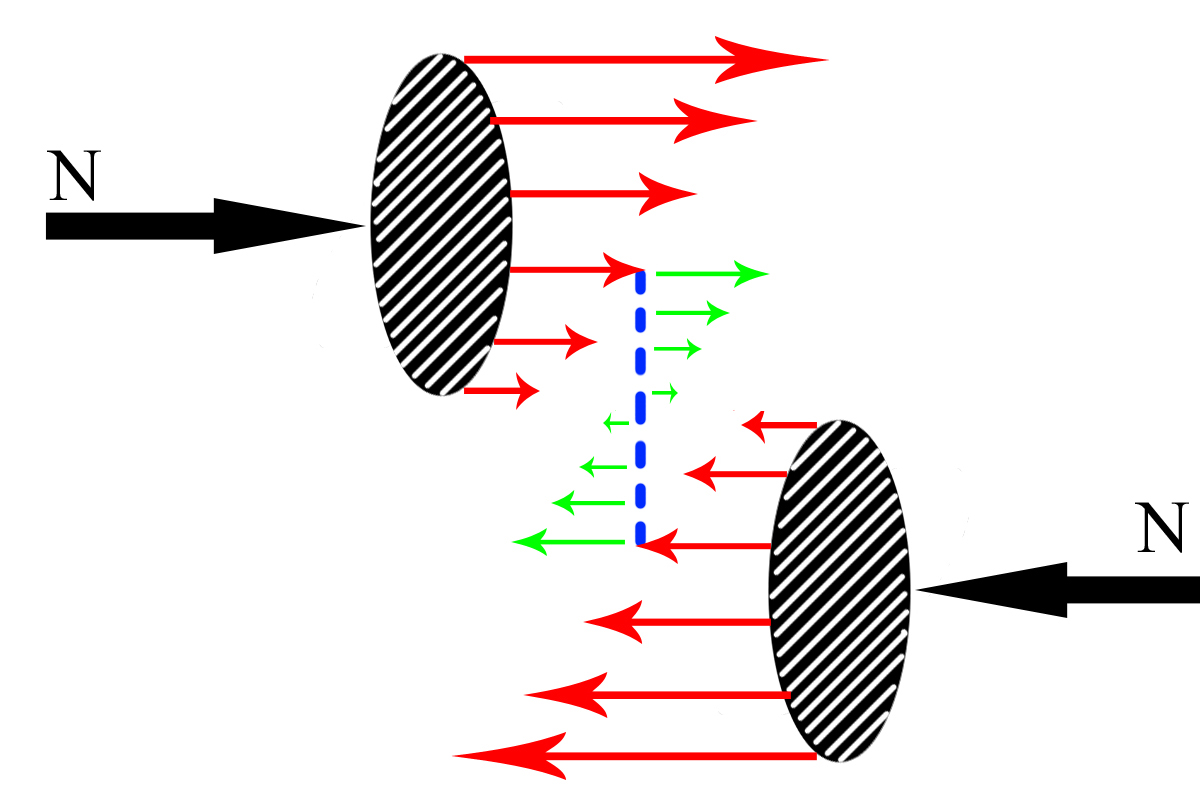}
\caption{\label{fig} Two different sources of hadroproduction: red arrows - particles produced by the preexisted partons, green - particles produced via the Pomeron exchange.}
\end{figure}

Therefore, the hadroproduction process in baryon-baryon high energy interactions could be decomposed into at least two distinct parts. These parts are characterized by two different sources of produced hadrons. The first one is associated with the baryon valence quarks and a quark-gluon cloud coupled to the valence quarks. Those partons preexist long time before the interaction and could be considered as being a thermalized statistical ensemble. When a coherence of these partonic systems is destroyed via strong interaction between the two colliding baryons these partons hadronize into particles released from the collision. The hadrons from this source are distributed presumably according to the Boltzmann-like exponential statistical distribution in transverse plane w.r.t. the interaction axis. The second source of hadrons is directly related to the virtual partons exchanged between two colliding partonic systems. In QCD this mechanism is described by the BFKL Pomeron exchange. The radiated partons from this Pomeron have presumably a typical for the pQCD power-law spectrum. Schematically figure~\ref{fig}  shows these two sources of particles produced in high energy baryonic collisions. This explanation is qualitative, however. 
This simple model is naive, though it allows to make a number of predictions which have been checked experimentally~\cite{OUR3,OUR4,OUR5}.

In this paper an attempt to study hadroproduction in heavy-ion collisions according to the introduced model is taken. In these collisions, due to a large number of colliding partons, extremely high energy densities, comparing to those in $pp$-collisions, can be obtained.
The experimental data measured in AuAu collisions at $\sqrt{s} = 200$ GeV/N by PHENIX~\cite{Adler:2003cb} and PbPb collisions at $\sqrt{s} = 2.76$ TeV/N by ALICE~\cite{Abelev:2012hxa} are considered here. 
Since the centre-of-mass energies per nucleon in these experiments are varied by a factor $\approx14$, a unified approach considering the energy density is suggested.
The energy density in heavy ion interactions is known to depend not only on the centre-of-mass energy, but also on the centrality of the collision. Hence, while the energy densities that can be reached at RHIC~\cite{Adler:2003cb}  and at LHC~\cite{Abelev:2012hxa} differ significantly, the energy density  in central collisions at RHIC might be of the same order as that in peripheral collisions at LHC.

Therefore, we use a simple parameterization ~\cite{Mishustin:2001ib} for the initial energy density which is motivated by several model calculations.
\begin{equation}
\label{eq:ed}
\varepsilon = \varepsilon_0 (\frac{s}{s_0})^{\alpha/2}{N_{coll}}^{\beta},
\end{equation}
where $\varepsilon_0 = 1$GeV/fm$^3$, $\alpha \approx 0.3$, $\beta \approx 0.5$ and $\sqrt{s_0} = 200$ GeV ~\cite{Mishustin:2001ib}. Here the second factor is responsible for the incident energy dependence, $\sqrt{s}$ is the c.m. collision energy, and the third one shows the dependence on the number of binary parton-parton collisions $N_{coll}$ which is related to the centrality of the collision.

\begin{figure}[!ht]
\includegraphics[width =9cm]{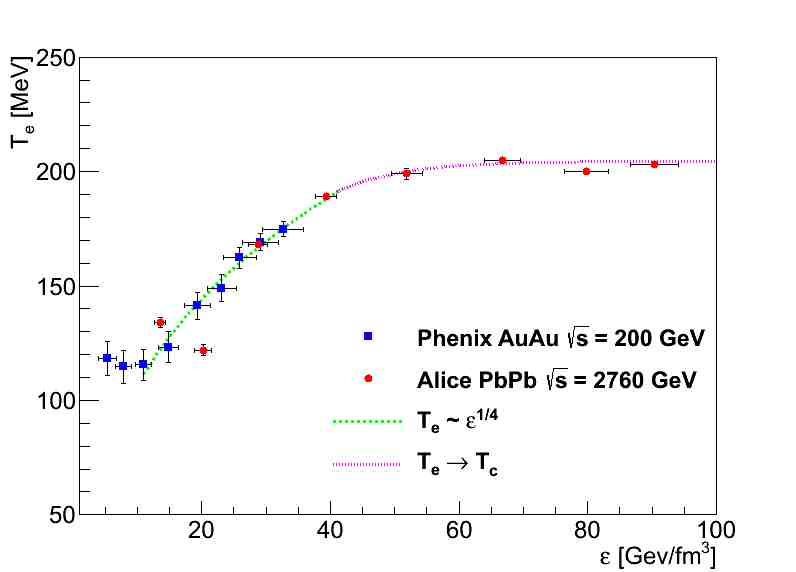}
\caption{\label{Tee} Temperature of charged particles released in heavy-ion collisions as function of energy density.}
\end{figure}

According to the introduced model, the exponential term stands for the radiation of thermalized particles with distributions similar to Boltzmann-like thermodynamics.
Fitting the introduced formula (eq.~\ref{eq:exppl}) to the measured experimental data allows to extract only the Boltzmann-like statistical distribution from the whole statistical ensemble of produced hadrons. This exponential distribution is characterized by a parameter $T_e$ analogous to the temperature in classical thermodynamics, which value is obtained from the fit. Therefore, it is interesting to study, how this temperature $T_e$ vary with the energy density obtained in the collision.

Figure ~\ref{Tee} shows this temperature $T_e$ as function of energy density.
As it was expected, the energy density obtained in central collisions at RHIC is similar to those in peripherial collisions at LHC, therefore a smooth transition between these two experiments is observed.
One can notice the interesting behavior of the temperature  as function of energy density ($T_e\propto \varepsilon ^{1/4}$), which is in a good agreement with the Stefan-Boltzmann law. Another observation on the temperature of the final state particles is that for high energy densities $T_e$ reaches a certain limit. This can be explained from QGP theory that considers the phase transition temperature $T_{c}$ from QGP to hadrons: the system cools until it reaches the critical temperature, thus, the temperature of the final state should be always below $T_{c}$.  Indeed, for high values of $\varepsilon$ one can notice, that the oberved critical temperature is $T_{c}\approx 200$ MeV, that is similar to previous theoretical observations~\cite{Hagedorn:1983wk}. 

In conclusion, the experimental results on charged hadron production in
heavy-ion collisions obtained in the PHENIX and ALICE experiments have been analyzed in the framework of the new approach. This approach allows to extract a part of the whole statistical ensemble of produced hadrons, those described by the Boltzmann-like statistical distribution only. This exponential distribution is characterized by a parameter $T_e$ analogous to the temperature in classical thermodynamics. It is found that the parameter $T_e$ depends universally on the collision energy density rather than on the collision energy.
Finally, the observed behavior of $T_e$ is formally similar to the Stefan-Boltzmann law well known in classic thermodynamics.


\end{document}